\documentclass[useAMS,usenatbib]{mn2e}
\usepackage{graphicx}
\title[Parent Stars of Extrasolar Planets. XII]{Parent Stars of Extrasolar Planets. XII. Additional evidence for trends with vsini, condensation temperature, and chromospheric activity}
\author[G.\ Gonzalez]{Guillermo Gonzalez$^{1}$\\
$^{1}$Grove City College, Rockwell Hall, 100 Campus Drive, Grove City, PA 16127, USA\\
}

\begin{document}

\date{Accepted ??. Received ??; in original form ??}

\pagerange{\pageref{firstpage}--\pageref{lastpage}} \pubyear{??}

\maketitle

\label{firstpage}

\begin{abstract}
Several recent studies have reported differences in vsini, abundance-condensation temperature trends, and chromospheric activity between samples of stars with and without Doppler-detected planets. These findings have been disputed, and the status of these results remains uncertain. We evaluate these claims using additional published data and find support for all three.
\end{abstract}

\section{Introduction}

Stars with planets detected with the Doppler method (SWPs) have been shown to be more metal-rich \citep{gg97, san01, vf05} and more massive \citep{law03, jj10} as a group when compared to similar stars without detected planets (non-SWPs). Recently, SWP and non-SWPs have also been reported to differ in vsini \citep{gg08, gg10a}, abundance-condensation temperature (T$_{\rm c}$) trends \citep{mel09, ram09, ram10, gg10b}, Li abundance \citep{gg08, israel09, gg10a}, and chromospheric activity \citep{gg08}, but each of these findings has been disputed. 

\citet{alv10} did not find a significant difference in vsini between their samples of SWPs and non-SWPs. \citet{bau10} found the Li abundance distributions to be indistinguishable between their samples of solar analog SWPs and non-SWPs. \citet{gh10} do not find a significant difference in abundance-T$_{\rm c}$ trends between their samples of solar analog SWPs and non-SWPs. Finally, \citet{cm11} do not find a significant difference in chromospheric activity between their samples of SWPs and non-SWPs.

The purpose of the present study is to revisit these controversies using published data and the method of analysis described in our recent series of papers \citep{gg08, gg10a, gg10b}. The paper is organized as follows. In Section 2 we compare the vsini distributions between SWPs and non-SWPs. In Section 3 we examine abundance-T$_{\rm c}$ trends. We compare chromospheric activity in Section 4. We summarize our results in Section 5.

\section{vsini}

\citet{gg08} and \citet{gg10a} compared vsini values of SWPs and non-SWPs using two samples. One of the samples was based on the extensive dataset of \citet{vf05}, which remains the best data for comparing SWP and non-SWP vsini values. However, since we completed our most recent analysis using their data, many new exoplanets have been discovered. The presence of undiscovered planets among the non-SWPs has been a source of unavoidable systematic error, but it is one that is becoming less important as new planets are discovered.

The full sample of stars from \citet{vf05} contains 1040 dwarfs; at the time of the paper's publication 85 of these stars with T$_{\rm eff} > 5500$ K were known to host Doppler-detected planets. To form our subsample, we first calculated the absolute visual magnitudes using the recalibrated Hipparcos parallaxes and then excluded those stars with a parallax value less than 10 times the parallax error. Next, we excluded stars with T$_{\rm eff} <$ 5500 K and T$_{\rm eff} >$ 6450 K; this is slightly broader than the range we had used in \citet{gg10a}, 5550 to 6250 K. Our final sample contains 99 SWPs and 627 non-SWPs. This compares to 82 SWPs and 594 non-SWPs used in \citet{gg10a}.\footnote{Note, the number of SWPs and non-SWPs in the present study with $5550 <$ T$_{\rm eff} <$ 6250 K are 93 and 578, respectively.}

We applied our method of analysis described in \citet{gg08} and \citet{gg10a} to the present data. In brief, we calculated a weighted average difference between the vsini value of an SWP and the vsini values of all the comparison stars using the inverse square of the $\Delta_{\rm 1}$ index as the weight. The $\Delta_{\rm 1}$ index is a measure of the distance between two stars in T$_{\rm eff}$-log $g$-[Fe/H]-M$_{\rm V}$ parameter space. We have made one minor change to this procedure compared to our previous studies. Previously, we had used [Fe/H] as one of the parameters needed to calculate the $\Delta_{\rm 1}$ index. However, using [Fe/H] could lead to a systematic error when thick disk stars are in the sample, since they have a different value of [$\alpha$/Fe] compared to thin disk stars. When we are considering the possible dependence of planet formation on composition, [M/H] is a better index to use than [Fe/H] \citep{gg09}. Therefore, we have replaced [Fe/H] with [M/H] when calculating the $\Delta_{\rm 1}$ index.

We show in Figure 1a the bias-corrected weighted average vsini differences between the SWP and non-SWP samples ($\Delta$vsini). We corrected the $\Delta$vsini values for bias in the same way as described in \citet{gg10a}. Briefly, the method involves splitting the non-SWP sample into two subsamples. We then calculated $\Delta$vsini values from these subsamples in the same way as was done with the original SWP and non-SWP samples. Any trends in these $\Delta$vsini values with T$_{\rm eff}$ are considered biases. The results presented in Figure 1a resemble those in Figure 12 of \citet{gg10a}, which was also based on the data of \citet{vf05}. As in \citet{gg10a}, we subtracted the average linear (bias) trend from the non-SWP $\Delta$vsini-T$_{\rm eff}$ data from the SWP $\Delta$vsini values. However, as is evident in Figure 11 of \citet{gg10a}, the required bias correction is not quite linear with T$_{\rm eff}$. 

\begin{figure}
  \includegraphics[width=3.5in]{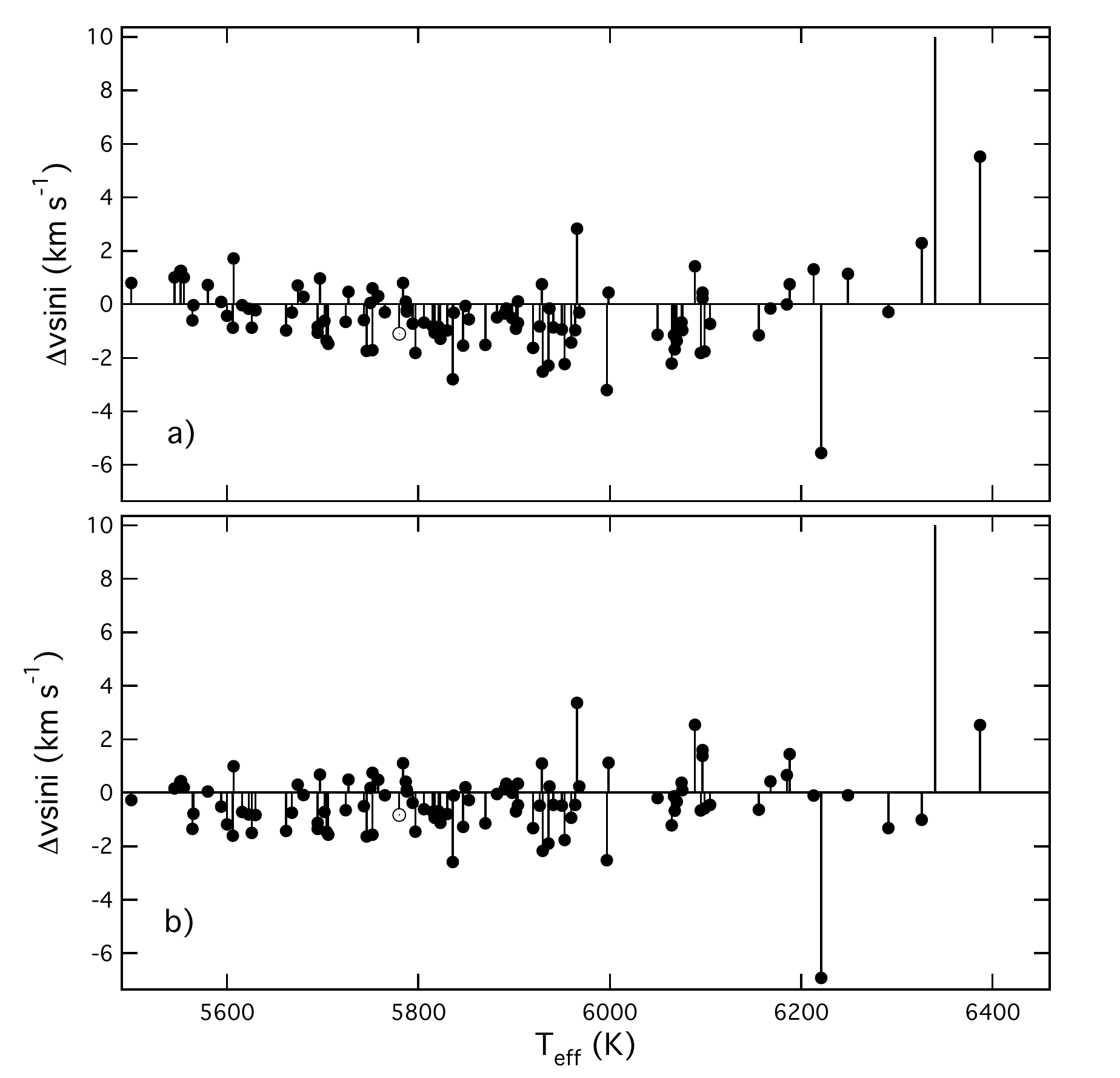}
 \caption{Weighted-average bias corrected vsini differences between SWPs and non-SWPs ($\Delta$vsini) using the samples described in the text. The open circle with the dot represents the solar value. One SWP with a $\Delta$vsini value of 16.4 km s$^{\rm -1}$ in $\bf{a}$ and 13.1 km s$^{\rm -1}$ in $\bf{b}$ is off the plotted range. A different bias correction was applied in $\bf{b}$; see text for details.}
\end{figure}

For this reason, we also corrected for bias using a second method. We calculated the vsini difference values in increments of 100 K using the non-SWP sample and applied these offsets to the vsini differences between the SWPs and non-SWPs. We show the results of this approach in Figure 1b. The overall pattern is similar to that in Figure 1a, but the distribution of vsini differences is flatter between 5500 and 6000 K. The average $\Delta$vsini value for the data plotted in Figure 1b between these two temperatures is $-0.46 \pm 0.96$ (s.e.) $\pm~0.11$ (s.e.m.) km s$^{\rm -1}$. The corresponding average $\Delta$vsini value determined by \citet{gg10a} from the \citet{vf05} data is $-0.66 \pm 1.08$ (s.e.) $\pm~0.13$ (s.e.m.) km s$^{\rm -1}$ (for T$_{\rm eff} = 5550$ to 6000 K).

\section{Trends with T$_{\rm c}$}

\citet{vf05} reported the abundances of five elements: Na, Si, Ti, Fe and Ni. While this is a small number of elements for our analysis, their T$_{\rm c}$ values span nearly the same range as the more extensive set of elements employed by \citet{gg10b}. In addition, the \citet{vf05} dataset is large and the abundances have small uncertainties. The samples of SWPs and non-SWPs we use in this sections are the same ones we used for the vsini analysis above.

We calculated the abundance-T$_{\rm c}$ slope for each star with simple linear least-squares. We then calculated the weighted-average [X/H]-T$_{\rm c}$ differences between the SWPs and non-SWPs using the same procedure as described above. However, in this case the bias corrections are small, allowing us to adjust the data with a simple linear fit, as in \citet{gg10b}. We show the corrected data in Figure 2. It resembles the data in Figure 2 of \citet{gg10b}.

\begin{figure}
  \includegraphics[width=3.5in]{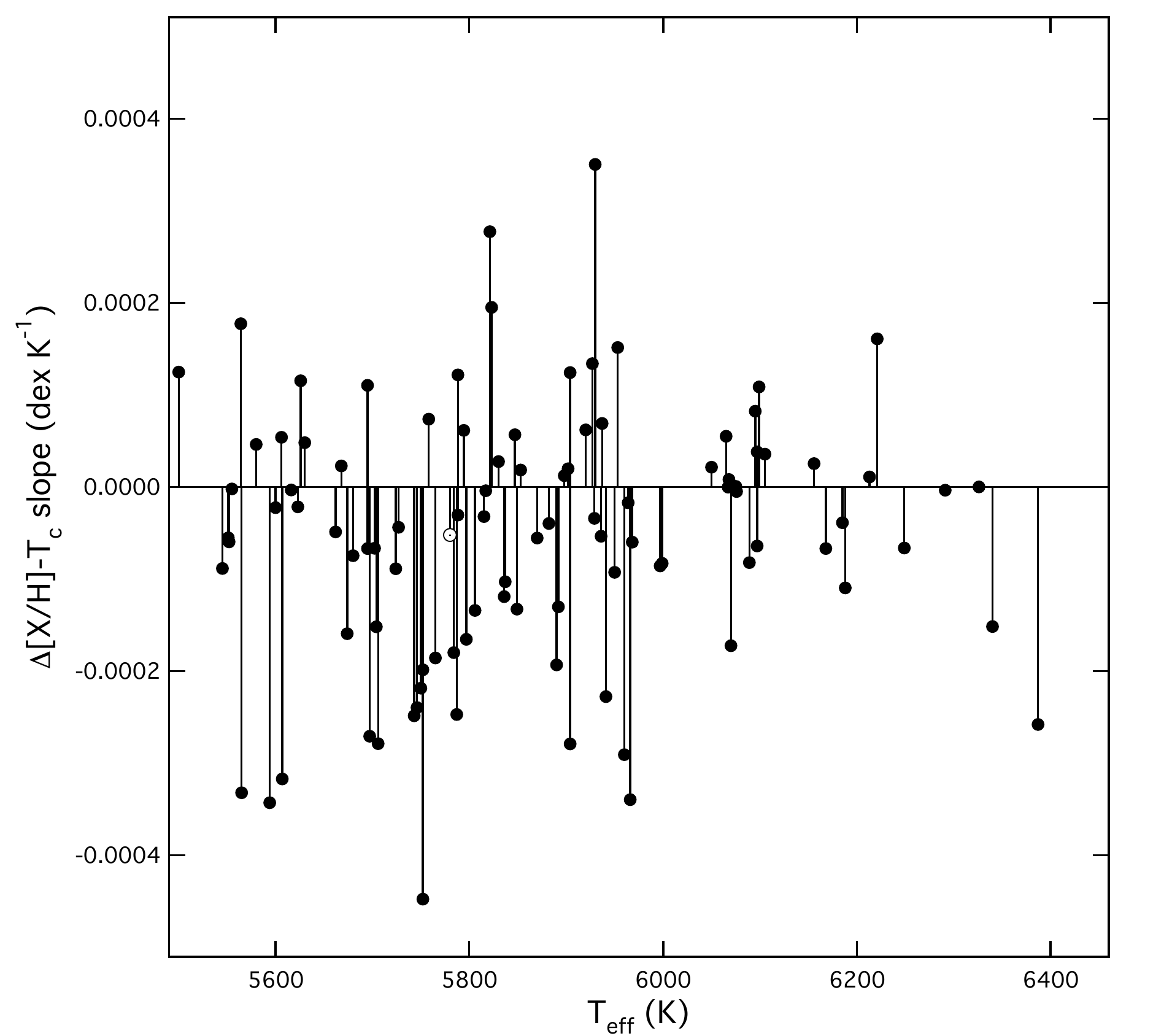}
 \caption{Weighted-average [X/H]-T$_{\rm c}$ slope differences between SWPs and non-SWPs using data from \citet{vf05}.}
\end{figure}

\citet{gg10b} also confirmed the discovery by \citet{mel09} that the more metal-rich SWPs display more negative [X/H]-T$_{\rm c}$ slopes than non-SWPs, while more metal-poor SWPs don't. We find very similar patterns in the current data, which we show in Figure 3.

\begin{figure}
  \includegraphics[width=3.5in]{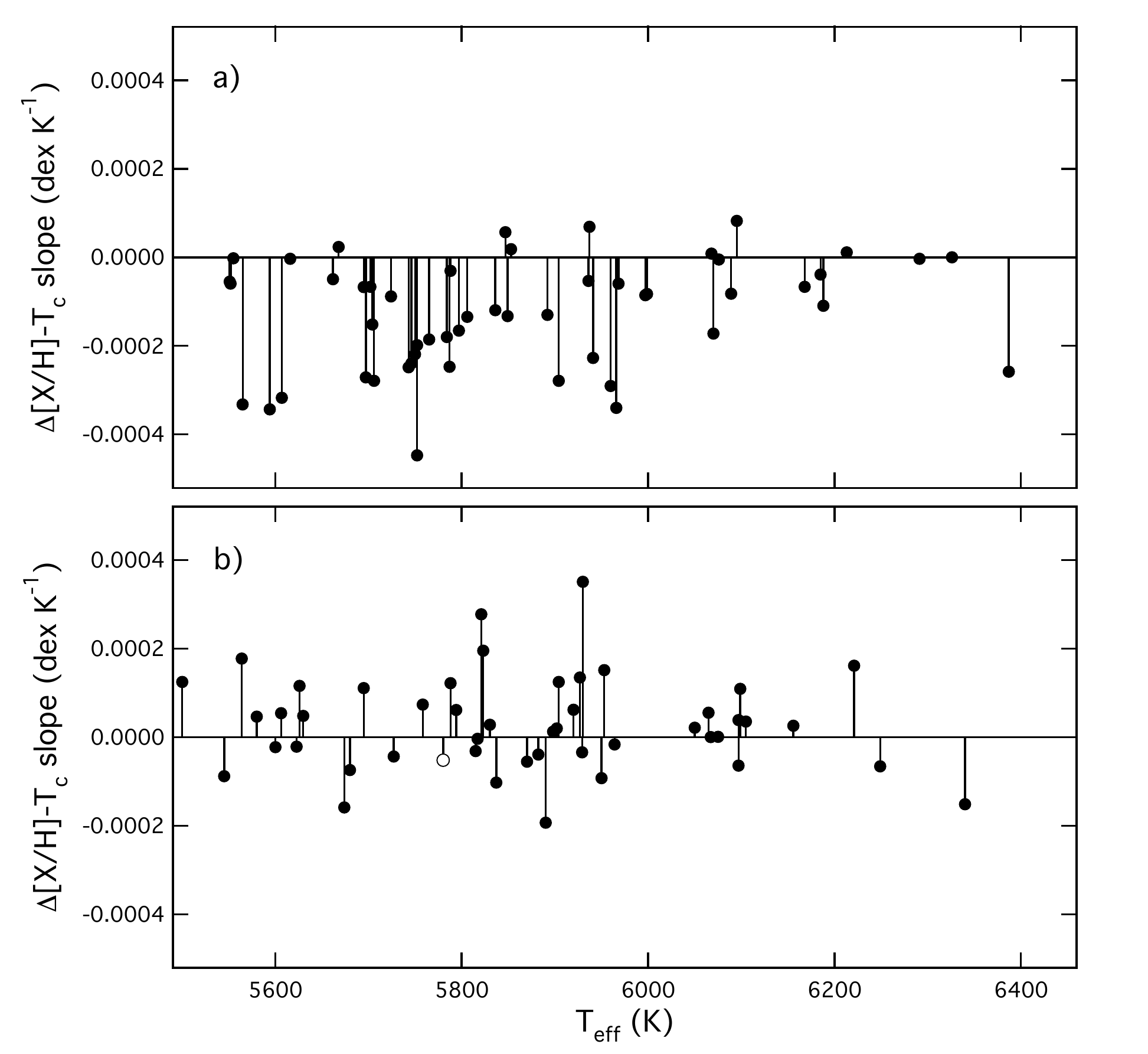}
 \caption{SWPs from Figure 2 with [M/H] $> 0.10$ are shown in $\bf{a}$, and SWPs with [M/H] $< 0.10$ are shown in $\bf{b}$.}
\end{figure}

\section{Chromospheric activity}

\citet{Isaac10} present chromospheric activity measurements of more than 2600 stars on the California Planet Search Program. In particular, they tabulate median values of $\log$ R$^{\rm \prime}_{\rm HK}$, which we use in the following analyses. We cross-referenced their data with the \citet{vf05} data and excluded stars with T$_{\rm eff} <$ 5500 K and T$_{\rm eff} >$ 6420 K; we also applied the same parallax quality cut used above in our vsini analysis. These cuts resulted in samples of 63 SWPs and 364 non-SWPs. We calculated weighted differences in $\log$ R$^{\rm \prime}_{\rm HK}$ using these two samples and corrected for bias using the same method used to produce Figure 1. We show the results in Figure 4. 

\begin{figure}
  \includegraphics[width=3.5in]{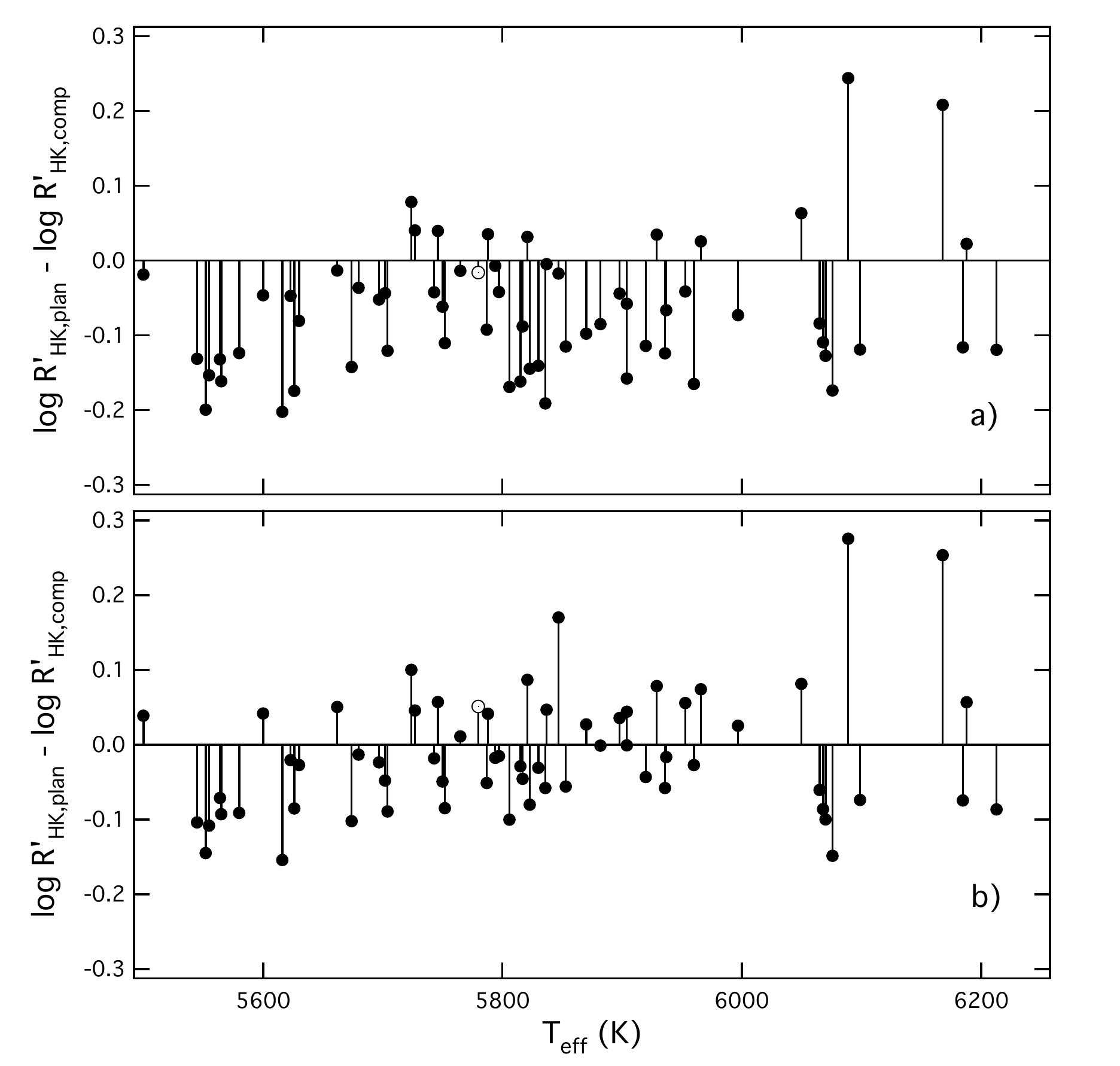}
 \caption{Weighted-average bias-corrected $\log$ R$^{\rm \prime}_{\rm HK}$ differences between SWPs and non-SWPs from \citet{Isaac10} shown in $\bf{a}$. A different cut is applied to the non-SWPs comparison sample in $\bf{b}$; see text for details.}
\end{figure}

The differences between the SWPs and non-SWPs are readily apparent in Figure 4a. However, we should be cautious in how we interpret this result. The SWP $\log$ R$^{\rm \prime}_{\rm HK}$ values range from -5.117 to -4.610, while it ranges from -5.206 to -3.879 for the non-SWPs. The apparent negative differences in $\log$ R$^{\rm \prime}_{\rm HK}$ for the majority of SWPs, then, could be due, in part, to the excess number of active stars in the non-SWP sample. In order to determine if this difference in the two samples can fully account for the pattern in Figure 4a, we have produced a second non-SWP sample using a conservative upper cutoff of -4.60 for $\log$ R$^{\rm \prime}_{\rm HK}$, resulting in a sample size of 307 stars. We repeated the above analysis using the same SWP sample as for Figure 4a and the new, more conservative non-SWP sample; the results are shown in Figure 4b.

There are far fewer SWPs with negative mean differences in $\log$ R$^{\rm \prime}_{\rm HK}$ in Figure 4b compared to Figure 4a. However, a trend is still evident. A linear least-squares fit to the data yields a slope of $(1.33 \pm  0.58) \times 10^{\rm -4}$ K$^{\rm -1}$; the mean weighted difference is zero at T$_{\rm eff} = 5925 $ K. The Pearson correlation coefficient for the data in Figure 4b is 0.283. This translates into a 5 \% probability that the trend is due to chance alone.

The truth should lie somewhere between Figures 4a and 4b. While the more conservative cut in $\log$ R$^{\rm \prime}_{\rm HK}$ values employed in preparing Figure 4b excludes stars that are much more active than any SWPs in our sample, it probably also excludes some stars whose $\log$ R$^{\rm \prime}_{\rm HK}$ values would not prevent Doppler detection of planets. For instance, HD 22049 has a $\log$ R$^{\rm \prime}_{\rm HK}$ value near -4.5, 0.1 unit larger than our cutoff.

We don't know why our conclusions are different from those of \citet{cm11}, but we do note some differences with their study. First, our samples are very different; we employ a much larger sample of non-SWPs. Second, our method of analysis compares SWPs to non-SWPs with similar physical parameters, including age. The $\log$ R$^{\rm \prime}_{\rm HK}$ index is known to be sensitive to age. Perhaps the difference we uncovered between SWPs and non-SWPs is too subtle to detect with other statistical approaches.

\section{Conclusions}

Using an updated version of the method of analysis described in \citep{gg08, gg10a, gg10b}, we have verified that there are significant differences in vsini, abundance-T$_{\rm C}$ trends, and chromospheric activity between SWPs and non-SWPs. We employed high-quality data from the literature, taking into account new planets that have been discovered since the data were originally published. 

We have verified that SWPs have significantly smaller values of vsini, abundance-T$_{\rm C}$ slope, and R$^{\rm \prime}_{\rm HK}$ compared to otherwise similar non-SWPs. For the case of the abundance-T$_{\rm C}$ slope differences, we also verified that they are significant when comparing stars with [M/H] $>$ 0.10, but not for more metal-poor stars. It is also notable that all three parameters display the largest differences between the SWP and non-SWP samples for T$_{\rm eff}$ less than about 5900 K.

\section*{Acknowledgments}

We thank the anonymous reviewer for helpful suggestions. We acknowledge financial support from the Grove City College.

\bsp

\label{lastpage}

\end{document}